\documentclass[sigconf,9pt]{acmart}

\renewcommand\footnotetextcopyrightpermission[1]{}
\AtBeginDocument{%
}

\usepackage{amsmath}
\usepackage{amsfonts}
\usepackage{amssymb}
\usepackage{multirow}
\usepackage{booktabs}
\usepackage[ruled,linesnumbered]{algorithm2e}
\usepackage[normalem]{ulem}
\usepackage{balance}
\usepackage[shortlabels]{enumitem}
\usepackage{threeparttable}
\usepackage{url}
\usepackage{graphicx}
\usepackage{textcomp}
\usepackage{xcolor}
\usepackage{subcaption}
\usepackage{hyperref}
\usepackage[dvipsnames]{xcolor}

\newcommand{\MethodName}{GTAC}

\newcommand{\qadd}[1]{{\color{black}{#1}}}

\newcommand{\qdel}[1]{}

\newcommand{\cdel}[1]{}
\newcommand{\ddel}[1]{}

\newcommand{\gadd}[1]{{\color{black}{#1}}}
\newcommand{\gdel}[1]{}

\newcommand{\wadd}[1]{{\color{black}{#1}}}
\newcommand{\wjdel}[1]{}
\newcommand{\wdel}[1]{}

\newcommand{\ldel}[1]{}

\begin{document}

\title{GTAC: A \underline{G}enerative \underline{T}ransformer for \underline{A}pproximate \underline{C}ircuits}









\author{Jingxin Wang\textsuperscript{\dag}, Shitong Guo\textsuperscript{\dag}, Wenhui Liang, Ruicheng Dai, Ruogu Ding, Xin Ning, Weikang Qian\textsuperscript{*}}
\affiliation{%
  \institution{Global College, Shanghai Jiao Tong University, Shanghai, China}
  \country{} 
}
\email{{jingxin.wang, jason.guo, qianwk}@sjtu.edu.cn}

\thanks{\textsuperscript{\dag} Co-first authors. \textsuperscript{*} Corresponding author.}

\begin{abstract}
Targeting error-tolerant applications, approximate computing relaxes rigid functional equivalence to significantly improve power, performance, and area\qdel{ (PPA)}. Traditional approximate logic synthesis (ALS) relies on incremental rewriting, limiting design space exploration.
\qadd{Meanwhile, the inherently probabilistic nature of Trans-former-based generative AI makes it a natural fit for generating approximate circuits.}
Exploiting this, we propose \MethodName{}, \qdel{the first}\qadd{an} end-to-end framework for arbitrary-scale generative ALS.
\qadd{To overcome the memory bottleneck of generative AI, }\MethodName{} \qdel{overcomes memory limits by decomposing}\qadd{partitions a} large circuit\qdel{s} into tractable \qdel{subgraphs}\qadd{subcircuits},
\qadd{applies a generative core to produce approximate candidates for each subcircuit, and finally selects proper candidates to form the final design.}
Its core generative Transformer utilizes a novel \qdel{Dynamic Referential DFS Encoding}\qadd{irredundant encoding} to \qdel{eliminate DAG structural redundancy}\qadd{compactly encode a circuit}, alongside \qdel{an error-tolerant}\qadd{a} masking mechanism \qdel{embedding \wdel{user-specified }error rate bounds directly into decoding}\qadd{to exclude designs violating the given error bound}.
Empowered by a self-evolutionary training \qdel{framework}\qadd{strategy}, \MethodName{} establishes a new paradigm \qdel{by demonstrating}\qadd{that demonstrates} superior performance\qdel{ (reducing}\qadd{: It reduces} delay by 30.9\% and gate count by 50.5\% over exact generative baselines\qdel{,} and \qdel{saving}\qadd{saves} 6.5\% area with a 4.3$\times$ speedup against traditional\qadd{ ALS} methods\qdel{), robust generalization across diverse logic structures, and unprecedented scalability on massive end-to-end designs}\wadd{. Furthermore, its irredundant encoding achieves a 33.3$\times$ reduction in sequence length and a 61.6$\times$ reduction in peak memory compared to conventional memoryless traversal.}
\end{abstract}


\maketitle

\vspace{-10pt}

\section{Introduction}\label{sect_intr}
\vspace{-1px}

\qdel{The relentless advancement of integrated circuits, fueled by Moore's \qdel{Law}\qadd{law} and the surge in AI and IoT applications, demands groundbreaking innovations in electronic design automation (EDA).
An important step in the EDA flow is logic synthesis.}%
\qadd{Logic synthesis is an important step in electronic design automation (EDA).}
\qdel{Within conventional EDA flows,}\qadd{Traditional} logic synthesis \qdel{prioritizes logical equivalence to guarantee}\qadd{always ensures} functional correctness~\cite{mishchenko2006improvements}.
Yet, this rigid requirement often \qdel{yields suboptimal outcomes in}\qadd{limits further improvement in} power, performance, and area (PPA)\qdel{, as it restricts the design space to precise mappings, curtailing optimization potential in resource-limited settings}.
\qadd{Targeting error-tolerant applications like image processing and data analytics~\cite{wang2024multi, shi2025alignment, wangunicircuit, armeniakos2022approximate, panteleaki2025leveraging}, approximate computing relaxes the rigid requirement on functional equivalence by allowing controlled inaccuracy to significantly improve PPA.}

Approximate logic synthesis (ALS) \ddel{extends approximate computing to logic-level optimization by relaxing exact equivalence constraints}is a \qdel{logic-level optimization technique}\qadd{logic synthesis approach} for generating approximate circuits\qdel{ under a given error constraint, enabling approximate computing in hardware designs}.
\qadd{It takes a given error constraint and a target circuit as inputs and produces an optimized approximate circuit satisfying the error constraint.}
Early works \qdel{proposed by Shin and Gupta\ddel{,} }simplified \qadd{sum-of-products Boolean expressions}\qdel{Boolean networks} under \qdel{bounded }error constraints to reduce \qdel{area and delay}\qadd{literal count}~\cite{ALSfirst}.
Subsequent works proposed various techniques to simplify circuit netlists, including \qdel{substitute-and-simplify strategies}\qadd{similar signal replacement}~\cite{SASIMI}, \qadd{approximate rewriting of AND-inverter graph (}AIG\qadd{)}\qdel{ rewriting}~\cite{chandrasekharan2016approximation, barbareschi2022catalog}, and advanced heuristic or learning-based searches~\cite{ALSRAC, SEALS, AccALS, AppResub, ma2021approximate, lee2023approximate}.
However, these methods\qdel{ are} typically \qdel{post-synthesis transformations applied}\qadd{perform incremental rewritings} to input exact \qdel{circuits}\qadd{netlists}, limiting their exploration of \qdel{novel topologies}\qadd{significantly different structures}.

\qdel{In parallel}\qadd{Meanwhile}, AI-assisted EDA has revolutionized circuit design through generative models, reinforcement learning (RL), and neural-guided methods~\cite{liang2023circuitops, chhabria2024generative}. 
Notable examples include Circuit Transformer (CT) and its extensions~\cite{li2025circuit, li2024ctrw}, which utilize masking-based decoding and tree search, as well as methods like ShortCircuit~\cite{tsaras2024shortcircuit} that rely on autoregressive decoding. 

Despite these advances, existing generative \qdel{Transformer}\qadd{AI} methods for circuits face fundamental bottlenecks. 
First, they suffer from poor generalization: Models like CT are constrained to synthesizing small circuits with specific \qdel{input/output dimensions}\qadd{numbers of inputs and outputs} and cannot generalize to \qdel{arbitrary topologies}\qadd{circuits with different numbers of inputs and outputs}. 
Second, their sequence-based representations cause severe memory explosion\qdel{ (out-of-memory)} when scaling up, making them incapable of \qdel{optimizing PPA for large-scale}\qadd{handling large} designs without a sophisticated partition-and-merge framework.
More importantly, \qdel{neural Transformers are}\qadd{generative AI is} inherently probabilistic, making strict functional exactness notoriously difficult to guarantee. 
However, this apparent flaw presents a unique opportunity\qadd{ for synthesizing approximate circuits}: \qdel{Transformers are \textit{naturally suited} for generating approximate circuits \qdel{within}\qadd{with} bounded errors.}%
By relaxing the rigid equivalence constraint, we can unleash the full \qdel{generative }power of \qadd{generative }AI to \qdel{explore novel topologies}\qadd{design high-quality approximate circuits}.
\qadd{While some recent efforts like GPTAC~\cite{yi2025gptac} have explored generative AI for producing approximate circuits, they remain restricted to small, domain-specific arithmetic blocks (\textit{e.g.}, multipliers) and lack a generalized, scalable framework capable of handling\qadd{ an} arbitrary\qdel{ large}-scale \qdel{logic designs}\qadd{circuits}.}


\qdel{Motivated by this insight}\qadd{To address the above issues}, we propose \MethodName{}~\footnote{To facilitate reproducibility and foster community adoption in AI-driven EDA, \MethodName{} is available at anonymous link: \href{https://anonymous.4open.science/r/GTAC}{https://anonymous.4open.science/r/GTAC}.},\qadd{ a \underline{g}enerative \underline{T}ransformer for \underline{a}pproximate \underline{c}ircuits. It is} a \qdel{pioneering }scalable framework\qdel{ that establishes a new \qdel{benchmark}\qadd{paradigm}} for \qdel{generative approximate logic synthesis}\qadd{generating approximate circuits}. 
To overcome the scalability \qdel{and generalization bottlenecks}\qadd{bottleneck}, \MethodName{} \qdel{operates as an end-to-end pipeline: it decomposes massive, arbitrary-scale}\qadd{first partitions a large} circuit\qdel{ graphs} into tractable \qdel{subgraphs}\qadd{subcircuits}, \qadd{then }generates \qdel{optimal}\qadd{optimized} approximate candidates \qdel{locally}\qadd{for each subcircuit}, and 
\qadd{finally selects proper candidates to form the final approximate circuit.}

\qdel{At the heart of this framework lies}\qadd{A key component of \MethodName{} is} a specialized generative Transformer engine. 
To conquer the memory explosion inherent in traditional sequence mapping, we introduce a novel \qdel{Dynamic Referential Depth-First Search (RefDFS) Encoding that strictly preserves \wadd{directed acyclic graph} (DAG) topology}\qadd{irredundant encoding to compactly encode a circuit}. 
Furthermore, to explicitly control the inherent randomness of the Transformer, we integrate \qdel{an error-tolerant}\qadd{a} masking mechanism that \qdel{embeds \qdel{user-defined logic}\qadd{the given error} constraint\qdel{s} directly into}\qadd{excludes designs violating the given error bound during} the autoregressive decoding process.
Our contributions are summarized as follows:

\begin{itemize}[leftmargin=*]

    \item \textbf{A Scalable \qdel{Paradigm}\qadd{Framework} for Generative ALS\qdel{ Framework}:}
    \qadd{\MethodName{} is a scalable framework for generating approximate circuits by leveraging a partition-and-merge pipeline.
    Compared to traditional ALS methods, it shifts from incremental netlist rewriting to direct circuit construction.}
    
    \item \textbf{A Specialized Generative Transformer Engine:} \qdel{Our novel architecture}\qadd{\MethodName{}} features \qdel{\wdel{Dynamic Referential DFS} RefDFS Encoding}\qadd{an irredundant encoding}, overcoming \qdel{sequence length}\qadd{memory} bottlenecks by eliminating \qdel{DAG structural}\qadd{the} redundancy\qadd{ in the existing circuit encoding}. 
    Additionally, \qdel{an error-tolerant}\qadd{a} masking mechanism seamlessly embeds user-\qdel{specified}\qadd{given error} constraint\qdel{s} into the token-level decoding process.
    
    \item \textbf{Self-Evolutionary Training Framework \& Superior PPA\qdel{ Trade-offs}:} \qadd{\MethodName{} performs a  self-evolutionary training guided}\qdel{Guided} by a PPA- and error-aware reward function, \qdel{our self-evolutionary training \wdel{strategy}\wadd{framework} }effectively \qdel{balances}\qadd{balancing} accuracy and hardware cost. 
    \qdel{\MethodName{}}\qadd{It} achieves highly competitive \qdel{qualitative }area reductions and significant\qdel{ evaluation} speedups over\qadd{ the} state-of-the-art ALS \qdel{baselines}\qadd{methods}.
    
\end{itemize}

\vspace{-0.5ex}
\section{Related Work}
\label{subsect_CGM}
\vspace{-0.5ex}

Recent generative circuit models, such as CT~\cite{li2025circuit, li2024ctrw} and ShortCircuit~\cite{tsaras2024shortcircuit}, excel at token-by-token generation but enforce strict functional equivalence, inherently limiting the explorable design space and potential PPA gains. 
\qdel{To address efficiency}\qadd{As a remedy}, models like GPTAC~\cite{yi2025gptac} introduce approximation via domain-specific pre-training for small arithmetic circuits. 
However, its heavy reliance on specific functional templates \qdel{prevents}\qadd{limits} structural adaptability, \qdel{halting}\qadd{hindering} its generalization to arbitrary logic topologies and scalability to massive designs.
\MethodName{} bridges these gaps. 
It adopts the decoding-with-constraints paradigm from exact models but \qdel{strategically }relaxes it \qdel{using explicit error thresholds}\qadd{by considering the given error bound}. 
Unlike GPTAC's domain-specific approach, \MethodName{} is entirely domain-agnostic, empowering large-scale, arbitrary logic approximation for superior PPA trade-offs.

\section{Preliminaries}\label{sect_prel}

This section introduces the ALS problem and \qdel{the circuit generative model}\qadd{a three-valued logic system}, establishing the foundation for \MethodName{}’s  framework.

\subsection{Approximate Logic Synthesis Problem}
\label{sec:approximate-equivalence-class}

Given a target Boolean function with $N$ inputs and $M$ outputs, $f: \mathbb{B}^N \to \mathbb{B}^M$\qdel{ (where $\mathbb{B} = \{0,1\}$)}, and a user-defined error bound $\epsilon$, the set of Boolean functions satisfying the given error bound is defined as:
\begin{equation}
\mathcal{C}_\epsilon(f) = \left\{ g \mid \mathcal{E}(g, f) \leq \epsilon \right\},
\end{equation}
where \(\mathcal{E}(g, f)\) is an error metric\qdel{a function} quantifying the deviation between an approximate function \(g\) and the target function \(f\).
A widely used error metric is the error rate (ER), calculated as:
\begin{equation}
\label{eq:er_metric}
\mathcal{E}(g, f) = \frac{1}{2^N} \sum_{\mathbf{x} \in \{0,1\}^N} \mathbb{I}\left(g(\mathbf{x}) \neq f(\mathbf{x})\right),
\end{equation}
where \(\mathbb{I}(\cdot)\) is the indicator function (1 if true and 0 otherwise).
ER measures the \qdel{fraction}\qadd{ratio} of\qadd{ the} input vectors yielding \qadd{incorrect }output\qdel{ mismatches}.
The objective of ALS is to identify a circuit in \(\mathcal{C}_\epsilon(f)\) that minimizes \qdel{resource metrics}\qadd{hardware cost}, such as area and delay.

\subsection{Three-Valued Logic}
\label{subsec:three_valued_logic}

\begin{figure}[htbp]
    \centering
    \includegraphics[width=1\linewidth]{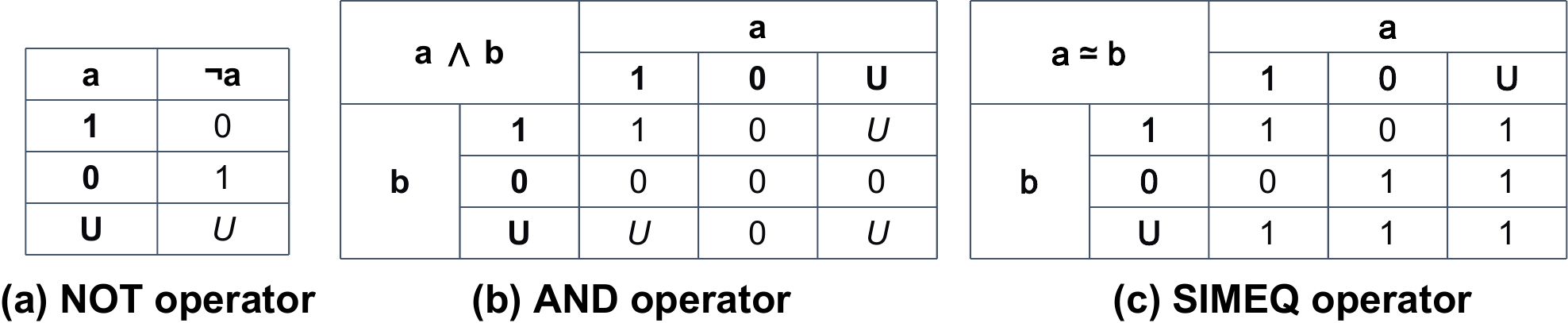}
    \caption{The truth tables for NOT, AND and SIMEQ operators in three-valued logic.}
    \label{fig:three_valued_logic}
    \vspace{-4mm}
\end{figure}

\MethodName{} is based on a three-valued logic system introduced in~\cite{li2025circuit}.
The system extends the standard Boolean domain $\{0, 1\}$ by adding an \emph{unknown} value, $U$.
Figs.~\ref{fig:three_valued_logic}(a) and (b) show the truth tables of the two operators in an AIG, NOT and AND, respectively, in this three-valued logic system.
\MethodName{} also utilizes a SIMEQ ($\simeq$) operator, with its truth table shown in Fig.~\ref{fig:three_valued_logic}(c). It acts as a relaxed equivalence check accommodating the unknown state (i.e., $U \simeq 0$ and $U \simeq 1$ are true). 


\begin{figure}
    \centering
    \includegraphics[width=0.95\linewidth]{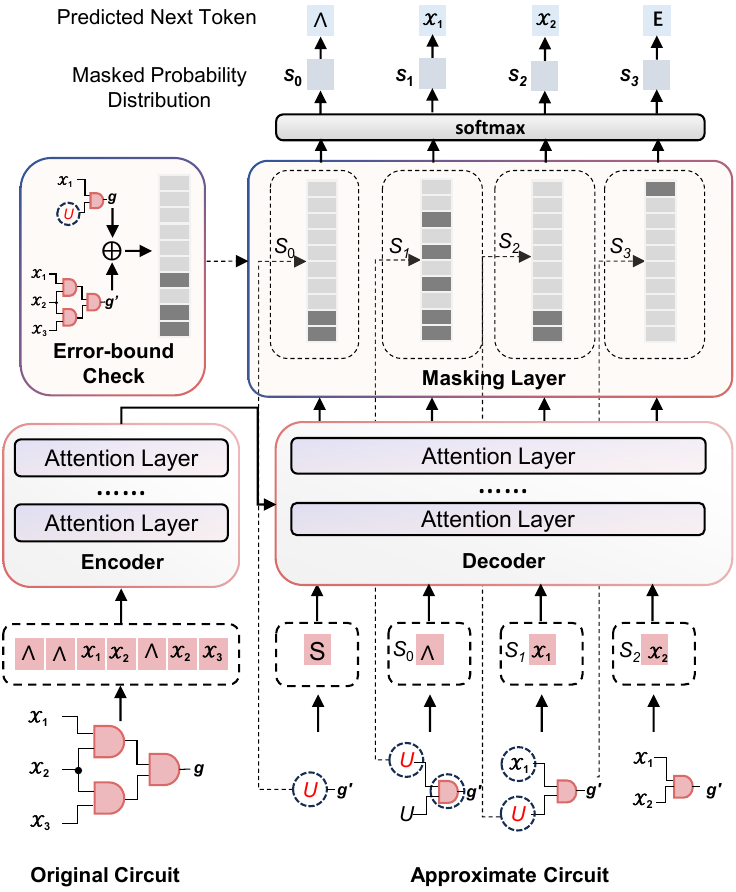}
    \caption{\qadd{The core of \MethodName{}, which also illustrates} the token-by-token inference process in \MethodName{}. 
    The encoder first processes the sequential representation of the original circuit. 
    During decoding, partial circuits are evaluated by the error-bound check engine, which guides the masking layer to dynamically filter out candidate tokens exceeding the error bound $\epsilon$, ensuring valid autoregressive prediction.
    \qadd{Each column vector corresponds to all tokens, where light and dark grey cells denote invalid and valid tokens, respectively.}
    \wdel{During decoding, the partial approximate circuit, utilizing 'U' placeholders for ungenerated logic gates, is dynamically evaluated by the Functional Logic Check. 
    This check computes the functional deviation and generates a hard mask via the Error-tolerant Masking Layer. 
    By modulating the probability distribution to filter out candidate tokens that exceed the user-specified error bound $\epsilon$, the model safely and autoregressively predicts the next valid structural token (e.g., $\Lambda$, $x_1$, $x_2$)}}.
    \label{fig:inference}
    \vspace{-8mm}
\end{figure}

\section{\MethodName{} Methodology}\label{sect_meth}


This section details the \MethodName{} methodology\qdel{ (overviewed in Figs.~\ref{fig:overview} and \ref{fig:gtac_pipeline})}.
We first \qdel{introduce the}\qadd{present its core, a} specialized generative Transformer \qdel{architecture, which processes AIG inputs while preserving DAG topology and utilizes an error-tolerant masking mechanism for bounded approximate decoding (Sec.~\ref{subsec:core_architecture})}\qadd{model, in Sec.~\ref{subsec:core_architecture}}.
To \qdel{optimize}\qadd{train} the model, we propose \qdel{a hybrid supervised and reinforcement learning strategy (Sec.~\ref{sec:training-strategy-and-loss-function}) coupled with }an iterative self-evolution \qdel{pipeline (Sec.~\ref{sec:iterative-self-improvement})}\qadd{strategy (Sec.~\ref{sec:training-strategy-and-loss-function})}.
Finally, to handle arbitrary-scale designs, we present an end-to-end scalable framework that partitions, infers, and globally composes subgraphs into a final approximate circuit (Sec.~\ref{subsec:scalable_framework}).

\subsection{\qdel{Model Architecture}\qadd{Core Generative Model of \MethodName{}}}
\label{subsec:core_architecture}

The core of \MethodName{}\qadd{, inspired by CT~\cite{li2025circuit},} is shown in Fig.~\ref{fig:inference}.
It takes an exact AIG and an ER bound as inputs and outputs an approximate AIG satisfying the ER bound.
It adopts the classical encoder-decoder Transformer architecture\qdel{ to map an exact AIG to an optimized approximate circuit}.
\begin{itemize}[leftmargin=*, nosep]
    \item \textbf{Encoder:}
    \qadd{The encoder takes a sequential representation of an exact AIG as an input and outputs contextual embeddings through a stack of attention layers. 
    \wdel{The encoder \qdel{is tasked with extracting}\qadd{extracts} the structural and functional context of the target exact circuit.}%
    To prevent the memory explosion typically caused by serializing large DAGs, \qdel{the input circuit is processed using our novel \emph{Dynamic Referential DFS (RefDFS) encoding} (Sec.~\ref{subsec:dyn_ref_dfs}), which explicitly eliminates structural redundancy}we propose an irredundant encoding to compactly encode a circuit, which will be detailed in Sec.~\ref{subsec:dyn_ref_dfs}}.
    
    \item \textbf{Decoder:} \qadd{The decoder takes the embeddings produced by the encoder as input and generates the approximate circuit step by step.
    As shown in the lower right of Fig.~\ref{fig:inference}, the generation process begins with the initial partial circuit consisting of a single $U$ node.
    At each step, it processes the encoder context and the partially generated token sequence, which encodes a partial circuit, to autoregressively predict the next circuit token.
    This corresponds to selecting a $U$ node in the circuit and replacing it by either a primary input or a logic gate (\textit{e.g.}, AND), whose own inputs are set as new $U$ nodes.
    To strictly enforce the given ER bound $\epsilon$ defined in our ALS formulation, a masking layer is inserted just before the decoder's final softmax layer.
    This mask prunes invalid tokens through an ER-bound check engine.
    The details of the ER-bound check engine and the masking mechanism will be described in Sections~\ref{subsec:scalable_error_evaluation} and~\ref{subsec:masking-mechanism}, respectively.}
\end{itemize}

\subsubsection{\qdel{Dynamic Referential DFS Encoding for Circuit Representation}\qadd{Irredundant Circuit Encoding}}
\label{subsec:dyn_ref_dfs}

\begin{figure}[!t]
    \centering
    \includegraphics[width=\columnwidth]{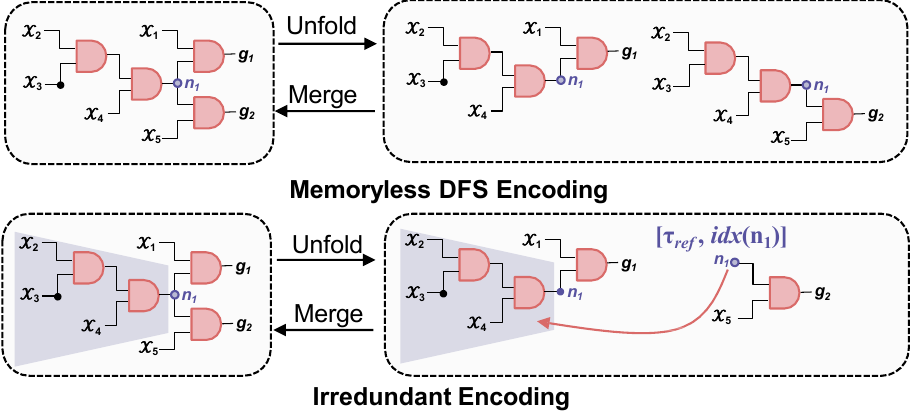}
    \caption{\gdel{Comparison of circuit encoding schemes.
Top (Conventional DFS): Memoryless DFS traversal unfolds a DAG into a tree, redundantly re-encoding multi-fanout nodes (e.g., $n_1$) in root sequences $s_{g1}$ and $s_{g2}$. 
In the worst case, this leads to an expanded tree size $N_{\text{tree}} = \mathcal{O}(F^D)$ for average fanin $F$ and depth $D$. 
Bottom (RefDFS): RefDFS replaces repeated expansions with a reference pair $[\tau_{\text{ref}}, \mathrm{idx}(n_1)]$, encoding the shared logic cone once. The resulting sequence preserves the original DAG topology and scales linearly with the number of unique nodes, i.e., $\Theta(|V|)$.}Comparison between memoryless DFS encoding and irredundant encoding.}
    \label{fig:refdfs_comparison}
    \vspace{-4mm}
\end{figure}

Transforming \qdel{DAG-structured circuits}\qadd{a circuit represented as a directed acyclic graph (DAG)} into\qadd{ a} sequential representation\qdel{s} is essential for Transformer-based processing.
To achieve this, \qdel{the work}\qadd{CT} proposes to apply a \emph{memoryless depth-first search (DFS)} to convert a DAG into a sequence~\cite{li2025circuit}.
\qadd{During the DFS, each time a node is visited, the DFS will be recursively applied to the first fanin of the node and then to the second fanin.}
Compared to traditional DFS, the memoryless DFS does not include an entry to record if a node is visited, meaning that it
\gdel{Conventional memoryless \gdel{depth-first search (DFS)}\gadd{DFS}~\cite{li2025circuit}}implicitly unfolds a \qdel{directed acyclic graph (DAG)}\qadd{DAG} into a tree (Fig.~\ref{fig:refdfs_comparison}), redundantly revisiting multi-fanout nodes along different traversal branches. 
This structural duplication enlarges the serialized representation, resulting in significantly higher memory usage\qdel{ due to the quadratic scaling of the self-attention mechanism of Transformer}\qadd{, as it grows quadratically} with\qadd{ the} sequence length\qadd{ for Transformer}.
\begin{example}
\label{exp:memoryless-DFS}
Consider the circuit in Fig.~\ref{fig:refdfs_comparison}, where node $n_1$ has two fanouts.
Under memoryless DFS, all nodes in the subgraph rooted at $n_1$ are visited twice along different branches, yielding the sequence \texttt{[AND, x1, AND, AND, x2, x3, x4, AND, AND, AND, x2, x3, x4, x5]}.
The repeated fragment \texttt{[AND, AND, x2, x3, x4]} appears because the same shared subgraph, \textit{i.e.}, the logic cone of $n_1$, is visited twice.
\end{example}
To \qdel{overcome this scalability bottleneck}\qadd{reduce the memory usage}, we propose \qdel{RefDFS}\qadd{an \emph{irredundant encoding}}, which \qdel{explicitly preserves\qadd{ the} DAG topology to }eliminate\qadd{s} duplication.
\qdel{For}\qadd{Specifically, for} a given subgraph $G$, we construct \qdel{an instance-specific}\qadd{a} vocabulary $\mathcal{V}_G$\qadd{ for it as}
\begin{equation}
\label{eqn:volcab}
\mathcal{V}_G = T_{\text{op}} \cup T_{\text{cst}} \cup T_{\text{var}}(G) \cup \{\tau_{\text{ref}}\},
\end{equation}
where $T_{\text{op}}$ and $T_{\text{cst}}$ are \wdel{static }operator (\textit{e.g.}, $\{\text{AND}, \text{NAND}\}$) and constant (\textit{e.g.}, $\{0, 1\}$) sets, respectively, 
$T_{\text{var}}(G) = \bigcup_{x \in PI(G)} \{ x, \bar{x} \}$ contains the literals for primary inputs $PI(G)$, and $\tau_{\text{ref}}$ is a topological reference token. 


\qadd{To produce the sequence encoding a given AIG, we perform a DFS on the AIG and record whether a node has been visited.
When an operator node $u$ is visited for the first time, we add its corresponding operator into the sequence and record the position of the operator in the sequence, denoted as $\text{idx}(u)$.
A future visit of the same node $u$ 
outputs the a token pair $(\tau_{\text{ref}}, \mathrm{idx}(u))$ instead of re-expanding the logic cone rooted at node $u$.
When an input literal is visited, it is always output directly, not replaced by references.}
This reference-based reuse bounds the \qdel{serialization}\qadd{sequence} length by the number of unique nodes. 
Consequently, the \qdel{self-attention }memory complexity is reduced from quadratic in the unfolded tree size to quadratic in the true AIG size.

\begin{example}
Consider the same circuit in Fig.~\ref{fig:refdfs_comparison}.
\qdel{RefDFS}\qadd{Our proposed irredundant encoding} avoids redundancy in the encoding sequence.
Specifically, when $n_1$ is \qdel{first }visited\qadd{ for the first time} via the first fanout branch, \qadd{the sequence \texttt{[AND, AND, x2, x3, x4]} corresponding to }its\qdel{ full} logic cone\qdel{ \texttt{[AND, AND, x2, x3, x4]}} is \qdel{serialized}\qadd{produced}.
Crucially, the position \qadd{of the operator of $n_1$ in the sequence is recorded as $\mathrm{idx}(n) = 3$}.
When the traversal reaches $n_1$ again through its second fanout, the reference pair \texttt{($\tau_{\text{ref}}, 3$)} is output instead of \qdel{re-traversing}\qadd{revisiting the logic cone of} $n_1$.
Consequently, the final sequence is \texttt{[AND, x1, AND, AND, x2, x3, x4, AND, $\tau_{\text{ref}}$, 3, x5]}.
Compared to \qadd{Example~\ref{exp:memoryless-DFS}}, the sequence length is reduced from 14 to 11\qdel{ tokens} by replacing a 5-token redundant fragment with a 2-token reference.
\end{example}

During inference, the exact DAG is reconstructed on-the-fly.
\gdel{Guided by the \qdel{emitted }$\tau_{\text{ref}}$ tokens, the decoder maintains a hash table corresponding to $\mathcal{R}$.} 
Upon proposing a new node, it checks\qdel{ the table} for functionally equivalent existing nodes. 
Matches are immediately merged via reference redirection, directly recovering a highly compact approximate DAG from the generated sequence.

\subsubsection{\qadd{ER-bound Check}}
\label{subsec:scalable_error_evaluation}

\qadd{An important component of \MethodName{} is an
\emph{ER-bound check engine}, which checks whether a partial circuit generated during the decoding process satisfies the ER bound.

Suppose that the input exact circuit and the partial circuit under check both have $N$ inputs and $M$ outputs.
Let the set of $M$ output functions of the input exact circuit be $f = (f_1, \ldots, f_M)$ and that of the partial circuit be $g = (g_1, \ldots, g_M)$.
As some inputs of a partial circuit are set to value $U$, the output value of a function $g_i$ under some input combinations may also be value $U$.
To calculate the ER of the partial circuit under the above special situation, we utilize the SIMEQ operator defined in Sec.~\ref{subsec:three_valued_logic}.
Specifically, for an input combination $\mathbf{x} \in \{0,1\}^N$, we define $f(\mathbf{x})=g(\mathbf{x})$ if $f_i(\mathbf{x}) \simeq g_i(\mathbf{x})$ for all $1\le i \le M$, and $f(\mathbf{x}) \neq g(\mathbf{x})$ otherwise.
The rationale behind this is that if $g_i(\mathbf{x}) = U$, it is still possible to be equivalent to $f_i(\mathbf{x})$, no matter whether $f_i(\mathbf{x})$ is $0$ or $1$.
In contrast, 
$g_i(\mathbf{x})$ cannot be equal to $f_i(\mathbf{x})$ only when $g_i(\mathbf{x}) = 0$ and $f_i(\mathbf{x}) = 1$ or $g_i(\mathbf{x}) = 1$ and $f_i(\mathbf{x}) = 0$.
The ER-bound check engine calculates the ER by applying the above definition of $f(\mathbf{x}) \neq g(\mathbf{x})$ into Eq.~\eqref{eq:er_metric}.
Then, it checks whether the calculated ER satisfies the ER bound.


For efficiency concern, in practice, when the number of inputs $N$ is smaller than a threshold $N_t$, the ER-bound check engine directly  traverses all $2^N$ input combinations to compute the ER, as shown in Eq.~\eqref{eq:er_metric}.
Otherwise, it randomly samples a subset of input combinations $X_m \subset \mathbb{B}^{N}$ to efficiently computing a high-confidence statistical approximation to the ER.}

\subsubsection{\qadd{Masking Mechanism for Invalid Tokens}}
\label{subsec:masking-mechanism}

\qadd{During the decoding process of \MethodName{}, it applies a masking mechanism to prune invalid token.
This mechanism exploits the ER-bound check engine.
Specifically, by applying the check engine at step $t$ of the decoding process, we obtain a set  $S_t^\epsilon$ of valid tokens:
\begin{equation}
S_t^\epsilon = \left\{ s \in \mathcal{V}_G \mid \mathcal{E}(g^{(t)}(s), f) \leq \epsilon \right\},
\end{equation}
where $g^{(t)}(s)$ is the Boolean function of the partial circuit constructed by appending the token $s$ to the current token sequence $s_1, \dots, s_{t-1}$,
$\mathcal{V}_G$ is the vocabulary of all possible tokens shown in Eq.~\eqref{eqn:volcab}, and the inequality $\mathcal{E}(g^{(t)}(s), f) \leq \epsilon$ is verified by the ER-bound check engine.

To enforce only considering the tokens in the set $S_t^\epsilon$ within the neural generation, as depicted in the overall architecture (Fig.~\ref{fig:inference}), we modulate the Transformer's output logits $\mathbf{z}_t$ with a hard mask $\mathbf{m}_t$, re-calibrating the probability distribution as $P(s_t | s_{1:t-1}) = \text{Softmax}(\mathbf{z}_t + \mathbf{m}_t)$, where $\mathbf{m}_t$ is set to $-\infty$ for any token $s \notin S_t^\epsilon$ and $0$ otherwise. 
This elegantly transforms the discrete logic synthesis problem into a constrained probabilistic generation task, ensuring that every sampled trajectory remains strictly within the \qdel{valid}\qadd{given} ER bound.}

\vspace{-2mm}
\subsection{\qdel{Pretraining Model}\qadd{Self-Evolutionary Training}}
\label{sec:training-strategy-and-loss-function}

\begin{figure*}[!t]
  \centering
  \includegraphics[width=1.0\linewidth]{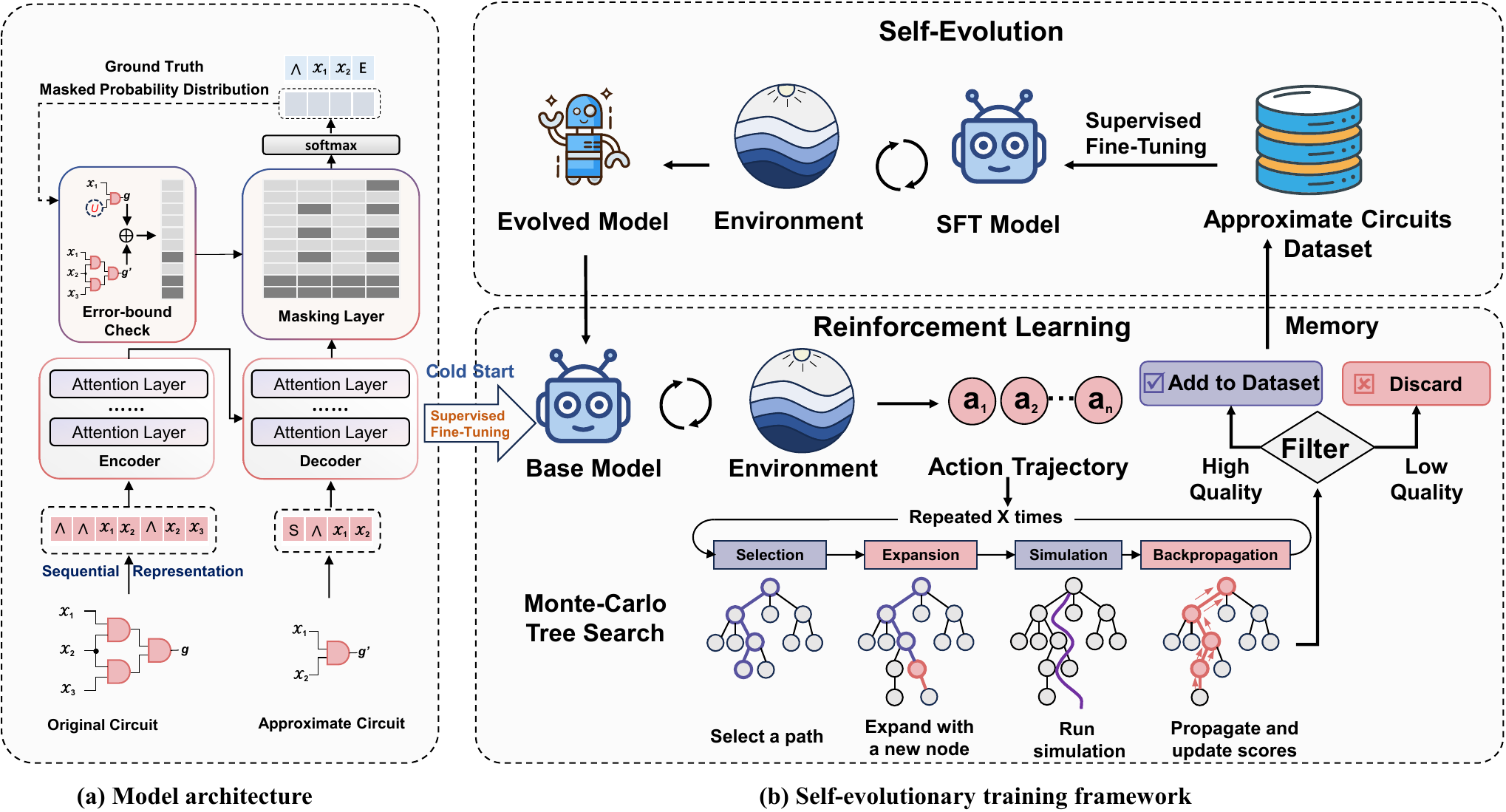}  
  \caption{Overview of the GTAC training framework. 
  (a) \textbf{Model architecture:} The generative Transformer takes the sequential representation of the original circuit and the partially generated approximate circuit to predict the next token during the training phase\qdel{ (where vectors denote token probabilities: \gdel{dark}\gadd{light} grey for error-masked invalid tokens, \gdel{light}\gadd{dark} grey for valid ones.)}.
  (b) \textbf{\wdel{Training phase and Self-evolution phase:}Self-evolutionary training framework:} The training pipeline begins with a cold start via supervised fine-tuning (SFT)\qadd{ followed by a reinforcement learning (RL) phase}.
  During the RL phase, Monte-Carlo tree search is used to explore action trajectories.
  Generated candidate circuits undergo a filtering process: High-quality pairs are added to the dataset\wdel{ memory}, iteratively driving the self-evolution of the model from a base SFT model to an evolved model.}
  \label{fig:overview}
  \vspace{-10pt}
\end{figure*}


To effectively train \MethodName{}, we propose a self-evolutionary training framework, as shown in Fig.~\ref{fig:overview}.
It integrates an offline supervised fine-tuning (SFT) initialization and an online RL exploration.
This paradigm ensures functional validity while providing a mechanism to minimize hardware costs (\textit{e.g.}, circuit size).

\subsubsection{Self-Evolution}
\label{sec:iterative-self-improvement}

To avoid the limitations of \wdel{purely imitating static data}\wadd{passively learning from fixed datasets} and continuously discover \qdel{novel}\qadd{new optimized} \qdel{Pareto-optimal topologies}\qadd{designs}, \MethodName{} employs an iterative self-evolution paradigm, as illustrated in the self-evolution block of Fig.~\ref{fig:overview}(b). 
This closed-loop pipeline seamlessly alternates between data generation and policy refinement. Specifically, in iteration $t$, the current model $\theta_{t-1}$, guided by Monte-Carlo tree search (MCTS), explores new token trajectories for a sampled batch of exact circuits $G_t$ to generate candidate approximate circuits $G'_t$.
These candidates are evaluated through a filter\qdel{;}\qadd{:} 
suboptimal circuits are discarded, while high-quality pairs\qadd{ of input exact circuit and its approximate counterpart} offering superior PPA are appended to the dataset memory.
Subsequently, the model is fine-tuned on this continuously expanding, high-quality memory pool to update its parameters to $\theta_t$. 
By progressively learning from its own filtered explorations rather than a static dataset, the model transcends its starting point, ultimately evolving into a highly capable policy.

\subsubsection{Supervised Fine-Tuning}

As illustrated in Fig.~\ref{fig:overview}(b), training begins with an SFT \emph{cold start} to initialize the base model, which then evolves through iterative updates on high-quality memory pools.
The \qdel{Base Model}\qadd{base model} undergoes \qdel{SFT}\qadd{a SFT} to learn the mapping from\qadd{ the} exact \qdel{circuit representations}\qadd{circuits} to their optimized approximate variants.
We optimize the model using the standard negative log-likelihood (cross-entropy) loss\qadd{ $\mathcal{L}_{\text{SFT}} = - \sum_{t=1}^T \log P_\theta(s_t^* \mid s_{1:t-1}^*)$} to maximize the probability of generating the ground-truth token sequences,
where $s_t^*$ represents the ground-truth token at step $t$, $s_{1:t-1}^*$ denotes the preceding ground-truth context, and $P_\theta$ is the conditional probability distribution parameterized by the Transformer model $\theta$. 
This stage effectively stabilizes the learning process, providing a robust\qdel{ behavioral} initialization before entering the exploration-heavy RL phase.
\subsubsection{Reinforcement Learning\wdel{ Fine-tuning}}

\qdel{Following}\qadd{After} the SFT cold start, we \qdel{introduce}\qadd{run} \qdel{a}\qadd{an} RL phase to steer the search toward \qdel{compact and functionally adequate }circuits\qadd{ with higher quality}. 
We formulate the generation process as a Markov \qdel{Decision Process}\qadd{decision process} (MDP), where \MethodName{} acts as the agent interacting with a circuit-building environment, as shown in Fig.~\ref{fig:overview}(b).
\wdel{This environment encapsulates a error-bound checker and a state tracker, evaluating proposed tokens and providing corresponding rewards.}%
This environment \qdel{simulates step-by-step circuit construction}\qadd{constructs the circuit token by token}\qdel{, utilizing a state tracker to maintain the partial \qdel{DAG topology}\qadd{circuit} and the error-bound check to estimate errors}.
For each generated token, it updates the circuit, evaluates its size and validity, and returns the corresponding reward.

\paragraph{State, Action, and Transition}
At time step $t$, the state $h_t$ corresponds to the partially generated circuit sequence up to the current step. 
The agent's action $a_t \in \mathcal{D}$ is the selection of the next token. The state transition is deterministic: appending action $a_t$ to the current sequence updates the state to $h_{t+1} = h_t \Vert a_t$, where $\Vert$ denotes the concatenation operator.

\paragraph{Reward Function}
The step reward $R(h_t, a_t)$ is designed to balance \qdel{circuit complexity and functional fidelity, comprising two main}\qadd{two} components\qadd{, circuit size and error}.
First, the \emph{size reward} $R_{\text{size}}$ discourages \qdel{excessive circuit complexity}\qadd{large circuit size} by penalizing the addition of new logic gates while \qdel{incentivizing structural}\qadd{encouraging gate} reuse. 
\qdel{Let $\mathcal{M}$ denote the event of a successful equivalent node merge. 
The size reward is formulated as}\qadd{It is calculated as}:
\begin{equation}
R_{\text{size}}(h_t, a_t) = \Delta(h_t, a_t) - \mathbb{I}\big(a_t \in \{\wedge, \bar{\wedge}\}\big),
\end{equation}
where $\mathbb{I}(\cdot)$ is the standard indicator function\qadd{ and $- \mathbb{I}\big(a_t \in \{\wedge, \bar{\wedge}\}\big)$ means that an immediate penalty of $-1$ is assigned whenever a new gate is added}.
The term $\Delta(h_t, a_t)$ \qdel{yields a positive reward if the chosen action $a_t$ triggers the merge event $\mathcal{M}$ under state $h_t$, and $0$ otherwise. 
An immediate penalty of $-1$ is assigned whenever a new fundamental logic gate is instantiated}\qadd{is the number of reused gates caused by the action  $a_t$ under the state $h_t$}.

Second, the \emph{error penalty} $R_{\text{error}}$ enforces the \qdel{functional approximation}\qadd{error} constraint.
Let $\mathcal{C}(h_{t+1})$ denote the\qdel{ instantiated} circuit parsed from the updated state sequence.
We penalize deviations from the user-specified error bound $\epsilon$ using a hinge-loss formulation~\cite{rosasco2004loss}:
\begin{equation}
R_{\text{error}}(h_{t+1}) = -\max\left(0, \mathcal{E}\big(\mathcal{C}(h_{t+1}), f\big) - \epsilon\right),
\end{equation}
where $\mathcal{E}(\mathcal{C}, f)$ measures \qdel{the empirical error of the generated circuit relative to the exact target function $f$}\qadd{the ER of the generated circuit $\mathcal{C}$}.
This effectively grants the model full exploratory freedom within the valid error \qdel{margin}\qadd{bound} while strictly penalizing any \qdel{boundary }violations.

The \qdel{overall}\qadd{total} step reward is defined as a weighted sum of \qdel{these}\qadd{the above two} components:
\begin{equation}
R(h_t, a_t) = \alpha \cdot R_{\text{size}}(h_t, a_t) + \beta \cdot R_{\text{error}}(h_{t+1}),
\end{equation}
where $\alpha$ and $\beta$ are hyperparameters controlling the trade-off between \qadd{circuit size} and error. 

\paragraph{Fine-tuning Objective}
To optimize the policy network $\pi_\theta$, we maximize the expected cumulative reward. 
For a generated trajectory $\tau = (h_1, a_1, \dots, h_T, a_T)$, let $G_t = \sum_{t'=t}^T R(h_{t'}, a_{t'})$ denote the un-discounted return from step $t$. The policy gradient loss $\mathcal{L}_{\text{RL}}$ is computed as: 
\begin{equation}
\mathcal{L}_{\text{RL}} = - \mathbb{E}_{\tau \sim \pi_\theta}\!\left[ \sum_{t=1}^T \log \pi_\theta(a_t \mid h_t) \cdot G_t \right].
\end{equation}
Minimizing this standard policy gradient objective efficiently steers the model parameters $\theta$ toward synthesizing circuits that are both minimal in size and strictly compliant with the error bound.

\paragraph{MCTS-Guided Local Inference}
\label{sec:inference-and-optimization}

To efficiently explore the design space for optimized sub-circuit topologies and explicitly characterize the \qdel{accuracy-compactness}\qadd{size-error} trade-off via Pareto optimization\wdel{(Sec.~\ref{sub: Pareto})}, we employ \wdel{Monte Carlo Tree Search (}MCTS to systematically explore the token-level solution space. 
As shown in Fig.~\ref{fig:overview}(b), to ensure that the MCTS specifically filters out invalid actions and only expands trajectories within the user-specified error bound $\epsilon$, we adopt MCTS guided by the predictor upper confidence bounds applied to trees (PUCT) rule~\cite{silver2017mastering}.\wdel{:
\begin{equation}
\text{PUCT}(a) = \frac{Q(a)}{N(a)} + c \cdot P(a) \cdot \sqrt{\frac{N(s)}{1 + N(a)}},
\end{equation}
where the state $s$ represents a partial token sequence, and the action $a$ is the next token prediction. 
$P(a)$ is the prior probability from the trained \MethodName{} Transformer. 
$Q(a)$ is the cumulative action value derived from the PPA- and error-aware reward function, while $N(s)$ and $N(a)$ are visit counts. 
$c=1$ is the exploration constant.}

As shown in Fig.~\ref{fig:inference}, each step of this token-by-token inference dynamically utilizes the masking layer to shape the prior probability $P(a)$. 
By exploring these diverse and strictly bounded token trajectories, \MethodName{} effectively samples a pool of optimized local candidates from the feasible set $\mathcal{C}_\epsilon(f)$.

\vspace{-3mm}
\subsection{Scalable Framework for Large Circuits}
\label{subsec:scalable_framework}

\begin{figure*}[t]
    \centering
    \includegraphics[width=1.0\textwidth]{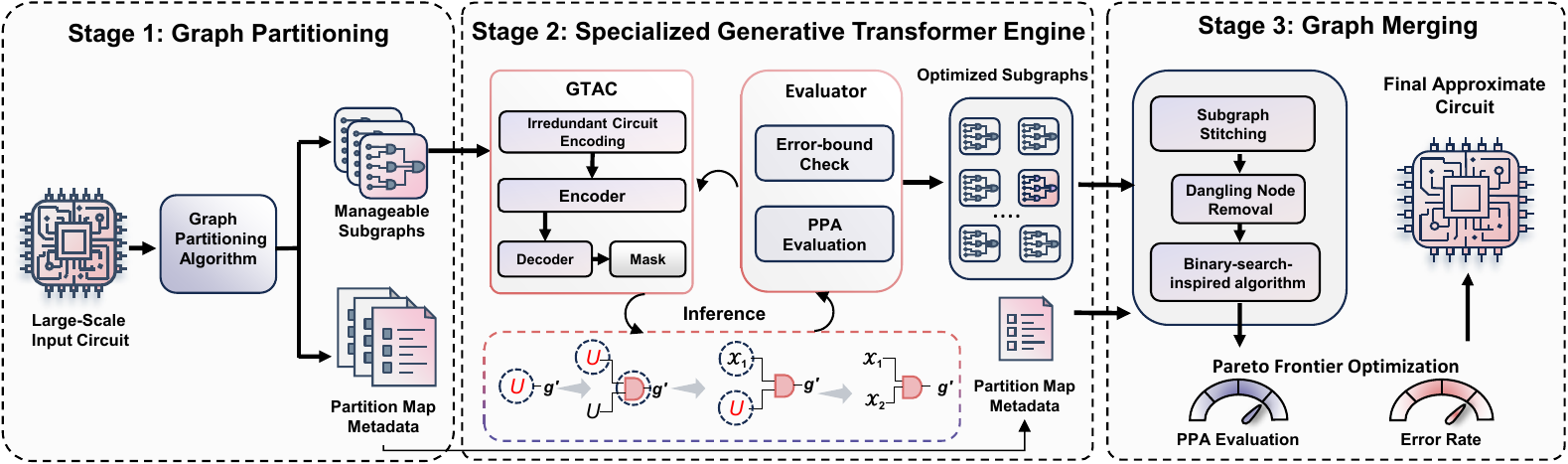} 
    \caption{End-to-end scalable pipeline for large-scale circuit approximation. 
    The framework operates through three sequential stages: (1) graph partitioning, which decomposes large input circuits into tractable subgraphs; 
    (2) the \MethodName{} core engine, which utilizes a Transformer model alongside the ER-bound check engine to dynamically generate local approximate candidates; 
    and (3) graph merging, which merges subgraph solutions into a final optimized approximate circuit while satisfying the ER bound.}
    \label{fig:gtac_pipeline}

\end{figure*}

As shown in Fig.~\ref{fig:gtac_pipeline}, we tackle large-scale designs through a divide-and-conquer pipeline by first partitioning the circuit into manageable subgraphs, then generating high-quality approximate candidates for each using the \MethodName{} core, and finally merging them into a complete design that satisfies the given ER bound.

\subsubsection{Graph Partitioning}
\label{subsec:dataset}

To handle a large-scale circuit, we first partition it into manageable subgraphs.
We employ an iterative expansion-based \qdel{partitioning }algorithm to extract subgraphs $G_i$.
In each iteration, the algorithm scans the nodes of the original AIG in\qadd{ a} topological order and selects the first node that has not yet been assigned to any subgraph to initialize a new subgraph.
The subgraph is then expanded by progressively including neighboring nodes in the \qdel{circuit graph}\qadd{AIG} while excluding nodes that have already been assigned to previously extracted subgraphs, ensuring non-overlapping partitions.  
The expansion stops once the subgraph size reaches a predefined bound $N_\textit{max}$.  
The\qadd{ entire} process repeats until all nodes are covered\qdel{, effectively avoiding the scalability bottlenecks of direct whole-circuit synthesis}.

\subsubsection{Local Candidate Generation}
\label{subsec:local_candidate_generation}

Each partitioned subgraph $G_i$ is processed independently by the \MethodName{} core engine.
\qadd{To provide a diverse pool of approximate candidates,
we sweep the ER bound $\epsilon \in \{0, 0.01, 0.05, 0.10\}$ to obtain a family of solutions balancing error and circuit size.
The resulting approximate candidate set, denoted as $\mathcal{C}_i$ for subgraph $G_i$, serves as a set of high-quality building blocks for the subsequent global composition.}

\subsubsection{Graph Merging}
\label{subsec:single_selection}

While the \MethodName{} core engine yields a set of high-quality local approximate candidates $\mathcal{C}_i$ for each subgraphs, \wdel{stitching them back is non-trivial; }the global composition space\qadd{ of size} $\mathcal{O}(\prod |\mathcal{C}_i|)$ remains prohibitively large.
To efficiently traverse this space, we employ a binary-search-inspired greedy algorithm\ldel{, a Pareto-driven greedy heuristic that incrementally refines circuit compositions along the \wdel{Pareto frontier of global PPA trade-offs}\wadd{global PPA Pareto frontier}} that iteratively constructs the final circuit.


Starting from the original exact circuit, each iteration evaluates the global ER induced by individually applying each approximate subgraph that has not yet been used, where applying an approximate subgraph means replacing the corresponding exact subgraph with it. 
\wadd{Specifically, this structural replacement involves stitching the optimized candidate back into the global topology and performing dangling node removal to clean up unreferenced logic gates.}
The subgraphs are ranked by their induced ERs, and the half with the smallest ERs are tentatively applied simultaneously. 
The resulting circuit is then evaluated for the global ER.
If the ER satisfies the target bound, the selected subgraphs are permanently applied, and the algorithm proceeds to the next iteration; otherwise, only the half with the smallest ERs is retained\ldel{ and the process repeats}, and if the resulting circuit still violates the ER bound, this set is further halved in the same manner until the bound is satisfied.
The iteration terminates when applying any remaining subgraph individually would violate the ER bound. 
This procedure incrementally integrates multiple approximate subgraphs while respecting the global ER bound, enabling efficient construction of high-quality approximate circuits without exhaustive search.

\vspace{-5mm}
\section{Experiment Results}\label{sect_res}

\subsection{Experimental Setup}

To comprehensively evaluate \MethodName{}, we adopt a multi-tiered experimental methodology. 
Initial baseline comparisons are conducted on the IWLS dataset to directly benchmark against state-of-the-art exact generative models. 
To further validate generalization and scalability, we expand our evaluation to diverse subgraphs and \qadd{the entire} large circuits from the EPFL~\cite{amaru2015epfl} and \wadd{Open}Cores~\cite{opencores} benchmark suites.

The \MethodName{} model employs an embedding dimension of 512, feedforward layers of size 2048, and a 12-layer encoder-decoder architecture with 8-head self-attention to effectively capture dependencies across circuit elements. 
We train \MethodName{} using the AdamW optimizer with a learning rate of $\eta=10^{-4}$ and a batch size of $64$ for 25 epochs on a single NVIDIA GeForce RTX 4090 GPU, which takes 20 hours.

\gdel{For enhanced assessment, particularly for arithmetic circuits, we consider two additional error metrics: Mean Relative Error Distance (MRED) and Mean Squared Error (MSE), defined as: $\text{MRED}(\qdel{\mathcal{G}}\qadd{g}, f) = \frac{1}{2^N} \sum_{\mathbf{x} \in \{0,1\}^N} \frac{|g(\mathbf{x}) - f(\mathbf{x})|}{\max(|f(\mathbf{x})|, 1)}$ and $\text{MSE}(\qdel{\mathcal{G}}\qadd{g}, f) = \frac{1}{2^N} \sum_{\mathbf{x} \in \{0,1\}^N} (g(\mathbf{x}) - f(\mathbf{x}))^2$, 
\qadd{where $g(\cdot)$ is the approximate function $g(\cdot)$ and $f(\cdot)$ is the exact reference function.}
In MRED, the denominator is defined as $\max(|f(\mathbf{x})|, 1)$ to avoid division by zero. Both metrics are highly relevant for applications where outputs are interpreted as numerical values. }

\vspace{-4mm}
\subsection{Evaluation of Core Generative Performance}

To ensure a fair and direct comparison with the state-of-the-art generative circuit model, Circuit Transformer~\cite{li2025circuit}, our initial baseline experiments are conducted on the same IWLS dataset used in~\cite{li2025circuit}. 

\subsubsection{Comparison with Existing Methods}

This section compares \MethodName{} against Circuit Transformer as well as state-of-the-art ALS methods HEDALS~\cite{10104162} and ALSRAC~\cite{ALSRAC}.
Circuit area and delay are measured after technology mapping using ABC~\cite{abc} with the Nangate 45 nm Open Cell Library~\cite{nangate45}.
The evaluation is performed on a dataset of 8K randomly generated circuits with average delay, area, and size measured by gate count as 84.47 ps, 22.35 $\mu m^2$, and 24.53, respectively. For ALS methods, the ER bound is set to 10\%.

As shown in Table~\ref{tab:comp_results}, \MethodName{} remains highly competitive against the prior ALS methods. Compared to ALSRAC, \MethodName{} presents improvements across all metrics (1.6\% in delay, 6.5\% in area, and 4.2\% in circuit size), with modest error increments. This is noteworthy as \MethodName{} was trained on \wdel{ALSRAC's data}\wadd{an approximate circuit dataset generated by ALSRAC~\cite{ALSRAC}}, indicating it has learned an optimization policy that surpasses its ``teacher''.

Crucially, \MethodName{} is also very efficient.
It has the lowest total runtime (9.47 min), being $2.8 \times$ to $4.3 \times$ faster than HEDALS and ALSRAC, respectively.
\wdel{The reason is because}\wadd{This efficiency stems from the fact that} traditional ALS heuristics are bottlenecked by sequential CPU execution, \wdel{but}\wadd{whereas} \MethodName{} leverages GPU-accelerated batch optimization to transform compute-bound synthesis into highly scalable tensor inference.
These results highlight \MethodName{}'s ability to balance accuracy and hardware cost.

\vspace{-3pt}
\begin{table}[htbp]
\centering
\small 
\caption{Performance comparison on averaged results from IWLS dataset. PPA and Error metrics are averaged across 8K testcases. (units: Delay (ps), Area ($\mu m^2$), Size (Gate count), Runtime (min)).}
\label{tab:comp_results}
\begin{tabular*}{\columnwidth}{@{\extracolsep{\fill}} l cccc c @{}}
\toprule
Methods & Delay$\downarrow$ & Area$\downarrow$ & Size$\downarrow$ & \gadd{ER}\gdel{MSE}$\downarrow$ & Runtime \\
\midrule
Circuit Transformer~\cite{li2025circuit} & 63.74 & 13.28 & 15.19 & 0.000 & 10.23 \\
HEDALS~\cite{10104162} & 43.43 & 6.52 & 7.80 & \gadd{0.076}\gdel{0.199} & 26.47 \\
ALSRAC~\cite{ALSRAC} & 44.76 & 6.43 & 7.85 & \gadd{0.078}\gdel{0.185} & 41.10 \\
\textbf{\MethodName{} (Ours)} & 44.05 & \textbf{6.01} & \textbf{7.52} & \gadd{0.082}\gdel{0.199} & \textbf{9.47} \\
\bottomrule
\end{tabular*}
\vspace{-4mm}
\end{table}

\subsubsection{Case Studies on IWLS Dataset Circuits}

\begin{table*}[htbp]
\centering
\caption{PPA and error metrics comparison of various methods across multiple cases.}
\label{tab:my_results}
\resizebox{\textwidth}{!}{%
\begin{tabular}{l cc ccc ccc ccc ccc}
\toprule
\multirow{2}{*}{Cases} & \multicolumn{2}{c}{\textbf{Circuit Transformer~\cite{li2025circuit}}} & \multicolumn{3}{c}{\textbf{ALSRAC~\cite{ALSRAC}}} & \multicolumn{3}{c}{\textbf{AppResub~\cite{AppResub}}} & \multicolumn{3}{c}{\textbf{HEDALS~\cite{10104162}}} & \multicolumn{3}{c}{\textbf{\MethodName{} (Ours)}} \\
\cmidrule(lr){2-3} \cmidrule(lr){4-6} \cmidrule(lr){7-9} \cmidrule(lr){10-12} \cmidrule(lr){13-15}
& Delay $\downarrow$ & Area $\downarrow$ & Delay $\downarrow$ & Area $\downarrow$ & \gdel{MRED}\gadd{ER} $\downarrow$ & Delay $\downarrow$ & Area $\downarrow$ & \gdel{MRED}\gadd{ER} $\downarrow$ & Delay $\downarrow$ & Area $\downarrow$ & \gdel{MRED}\gadd{ER} $\downarrow$ & Delay $\downarrow$ & Area $\downarrow$ & \gdel{MRED}\gadd{ER} $\downarrow$ \\
\midrule
Case 1  & 96.29 & 13.81 & 20.28 & 1.06  & 0.051 & 20.28 & 1.06 & 0.051 & 20.28 & 1.06 & 0.051 & 20.28 & 1.06 & 0.051 \\
Case 2  & 87.68 & 10.61 & 39.74 & 3.72  & 0.059 & 39.74 & 3.72 & 0.059 & 23.72 & 2.39 & 0.066 & 23.72 & 2.39 & 0.066 \\
Case 3  & 72.65 & 11.69 & 32.88 & 3.45  & 0.078 & 35.62 & 3.19 & 0.078 & 32.88 & 3.45 & 0.078 & 32.88 & 3.45 & 0.078 \\
Case 4  & 68.55 & 13.52 & 41.43 & 6.90  & 0.039 & 38.19 & 5.59 & 0.039 & 47.85 & 8.23 & 0.043 & 41.43 & \textbf{5.31} & 0.086 \\
Case 5  & 64.81 & 15.95 & 50.55 & 5.58  & 0.078 & 57.75 & 5.57 & 0.078 & 49.68 & 5.32 & 0.078 & \textbf{48.78} & 5.32 & 0.078 \\
Case 6  & 61.89 & 17.00 & 0.00  & 0.00  & 0.078 & 0.00  & 0.00  & 0.078 & 0.00  & 0.00 & 0.078 & 0.00  & 0.00  & 0.078 \\
Case 7  & 64.24 & 7.98  & 45.99 & 5.05  & 0.078 & 47.29 & 5.85 & 0.063 & 45.99 & 5.05 & 0.078 & \textbf{45.30} & \textbf{5.58} & 0.094 \\
Case 8  & 70.10 & 12.74 & 47.48 & 7.70  & 0.094 & 50.72 & 6.38 & 0.094 & 47.48 & 7.70 & 0.094 & 47.48 & 7.70  & 0.094 \\
Case 9  & 64.24 & 9.57  & 44.33 & 7.97  & 0.086 & 44.97 & 7.97 & 0.047 & 44.97 & 7.97 & 0.047 & 50.08 & \textbf{6.90} & 0.063 \\
Case 10 & 59.59 & 11.42 & 45.95 & 6.64  & 0.078 & 51.31 & 7.70 & 0.055 & 42.54 & 6.90 & 0.070 & 50.15 & \textbf{6.38} & 0.094 \\
\midrule
Average & 71.00 & 12.43 & 36.82 & 4.81  & 0.072 & 38.59 & 4.70 & 0.064 & 35.54 & 4.81 & 0.068 & 36.02 & \textbf{4.41} & 0.078 \\
\bottomrule
\end{tabular}%
}
\end{table*}

To further illustrate the behavior of different methods on individual circuits, we randomly selected 10 cases from the IWLS FFWs 2023 dataset~\cite{Mishchenko2023iwlscontest}.
Detailed results are reported in Table~\ref{tab:my_results}.

For half of the cases, \MethodName{} achieves smaller delay and/or area than AppResub~\cite{AppResub} and HEDALS~\cite{10104162}.
Notably, for Case~2, \MethodName{} matches the PPA of HEDALS, and reduces delay by 40\% and area by 36\% compared to AppResub, together with smaller errors.

We also find that Case 6 is special. It can be simplified to a constant output circuit within the error bound, leading both \MethodName{} and AppResub to eliminate the gates completely.
These results highlight \MethodName{}’s ability to consistently find Pareto-favorable trade-offs on a case-by-case basis.

\subsubsection{Pareto Front Analysis}
\label{sub: Pareto}

\begin{figure}
    \centering
    \includegraphics[width=1\linewidth]{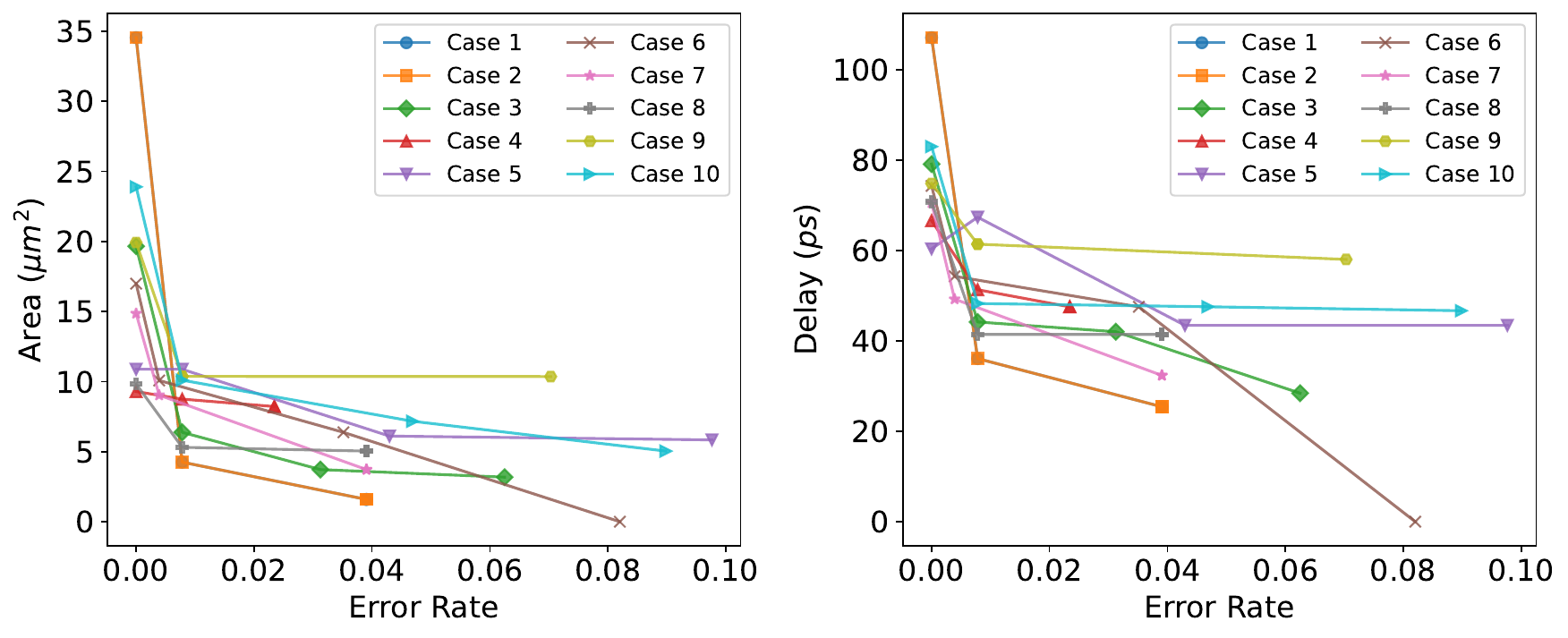}
    \caption{Pareto front of the evaluated design cases: (a) Error rate versus area; (b) Error rate versus delay.}
    \label{fig:pareto_cases}
    \vspace{-10pt}
\end{figure}

\begin{figure}
    \centering
    \includegraphics[width=1\linewidth]{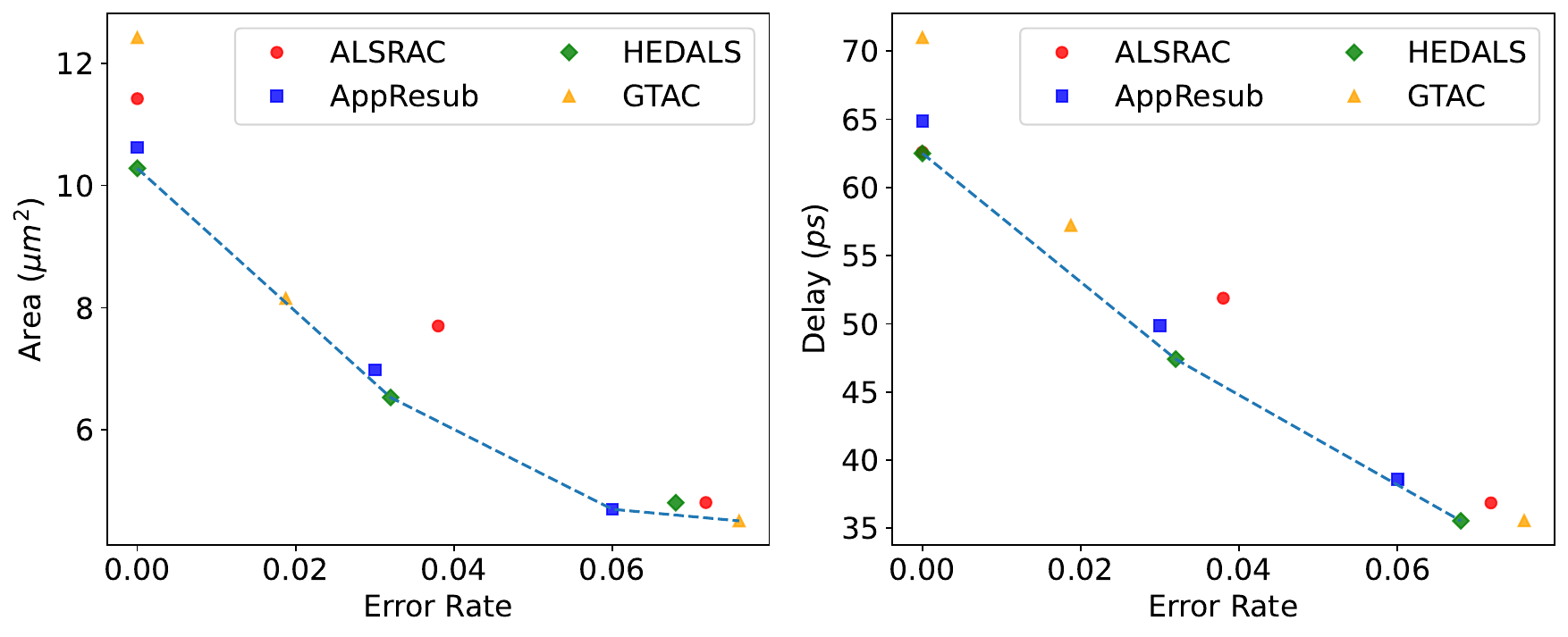}
    \caption{Pareto front comparison among ALSRAC, AppResub, HEDALS, and \MethodName{}: (a) Error rate versus area; (b) Error rate versus delay.}
    \label{fig:pareto_act}
    \vspace{-4mm}
\end{figure}

\wdel{As shown in Fig.~\ref{fig:pareto_cases} and Fig.~\ref{fig:pareto_act}, the Pareto fronts capture the trade-offs among error rate, delay, and area.}\wadd{To explicitly evaluate the capability of traversing the complex design space, we analyze the Pareto frontiers detailing the interplay between functional accuracy and hardware efficiency.} In Fig.~\ref{fig:pareto_cases}, the ten randomly sampled design cases demonstrate diverse Pareto-optimal sets, reflecting different trade-offs between accuracy and hardware cost.
In Fig.~\ref{fig:pareto_act}, the comparison shows that HEDALS and AppResub define the optimal PPA/error fronts in lower error bound regions, while \MethodName{} remains competitive against the prior ALS methods, and achieve\gadd{s} the lowest area against all methods when ER bound is 10\%.

\subsubsection{Self-Evolution on Approximate Datasets}

We self-evolve \MethodName{} using another approximate circuit dataset constructed based on the IWLS dataset. 
This dataset, featuring diverse circuit pairs with controlled error bounds, enables \MethodName{} to effectively balance performance and accuracy. 
Post-evolution, \MethodName{} achieves significant reductions of 21.6\% in delay, 33.5\% in area, and 30.8\% in size. 
These results underscore the synergistic impact of \MethodName{}'s self-evolution (Sec.~\ref{sec:iterative-self-improvement}), enhancing efficiency while maintaining acceptable error levels for approximate computing.

\subsubsection{Ablation Study on Key Components}

\wadd{To validate the essential components of \MethodName{}, including the masking mechanism, composite loss, and the approximate dataset, we conducted ablation studies under identical benchmark setups. 
The full \MethodName{} consistently achieves the best PPA-error trade-off. 
Notably, while removing the masking mechanism eliminates errors entirely (0.00 MRED), it incurs a severe PPA penalty: delay increases by 90\%, area by 267\%, and size by 222\%.
This highlights the significant PPA savings unlocked by our approximation strategy. 
Additionally, optimizing with only CE loss or excluding the approximate dataset degrades overall PPA by 15\% and 28\%, respectively. 
These results underscore that all components are highly synergistic and indispensable for maximizing PPA gains within acceptable error bounds.}
\subsection{\gadd{Scalability of \wdel{Proposed }Irredundant Encoding}}

\gadd{To analyze the scalability of \MethodName{}'s irredundant encoding, we compare sequence length, encoding runtime, and inference-time peak memory across representative EPFL benchmarks in Table~\ref{tab:encoding_comparison}. 
Peak memory is measured on GPU during the forward pass of \qdel{MultiHeadAttention}\qadd{the multi-head attention} (batch size 64, attention heads 8). 
Our proposed irredundant encoding yields substantially shorter sequences, shorter encoding time, and lower memory usage, whereas memoryless DFS results in out-of-memory on complex circuits.}
\wadd{Specifically, the irredundant encoding achieves an average reduction of 33.3$\times$ in sequence length, 21.7$\times$ in encoding time, and $61.6\times$ in peak memory.}





\subsection{Generalization Across Diverse Logic Structures}

To demonstrate \MethodName{}'s scalable generalization to arbitrary input/output (I/O) configurations beyond the training distribution, we evaluate it on structurally complex subgraphs. We deliberately select arithmetic (\texttt{multiplier}) and control-heavy (\texttt{arbiter}) circuits from EPFL and Core benchmarks; their dense, multi-fanout topologies serve as rigorous stress tests for structural adaptability and memory bounds. \gadd{The size bound $N_{max}$ for the partitioned subgraph is set to 50. Circuit area and delay are measured after technology mapping using ABC~\cite{abc} with ASAP7~\cite{asap7}.}

As shown in Table~\ref{tab:subgraph_case_study_combined}, the baseline CT consistently encounters Out-of-Memory (OOM) failures due to the quadratic memory explosion of conventional sequence encoding on complex DAGs. Conversely, empowered by our irredundant encoding, \MethodName{} bypasses this bottleneck entirely. 
It successfully generates optimized approximate candidates across all challenging cases, proving its robust memory efficiency and capability to synthesize \wadd{structurally diverse} logic without fixed I/O restrictions.

\begin{table}[t!]
\centering
\footnotesize
\setlength{\tabcolsep}{4pt} 
\caption{\wadd{Average encoding performance on EPFL subgraphs. Seq., Time (ms), and Mem (GB) denote sequence length, encoding runtime, and inference peak memory.}}
\label{tab:encoding_comparison}
\resizebox{\columnwidth}{!}{ 
\begin{tabular}{l ccc ccc ccc}
\toprule
\multirow{2}{*}{Benchmark}
& \multicolumn{3}{c}{Memoryless}
& \multicolumn{3}{c}{Irredundant (Ours)}
& \multicolumn{3}{c}{Reduction Ratio} \\

\cmidrule(lr){2-4}
\cmidrule(lr){5-7}
\cmidrule(lr){8-10}

& Seq.\ & Time & Mem
& Seq.\ & Time & Mem
& Seq.$\downarrow$ & Time$\downarrow$ & Mem$\downarrow$ \\

\midrule
adder      & 281.45   & 0.38 & 1.407 & 145.91 & 0.31 & 0.372 & 1.93$\times$ & 1.23$\times$ & 3.78$\times$ \\
arbiter    & 108.36   & 0.17 & 0.296 & 104.73 & 0.30 & 0.296 & 1.03$\times$ & 0.57$\times$ & 1$\times$ \\
multiplier & 13678.19 & 21.05& OOM   & 137.16 & 0.29 & 0.428 & 99.7$\times$ & 72.6$\times$ & $> 112.1\times$ \\
priority   & 3021.38  & 4.46 & OOM   & 124.95 & 0.28 & 0.371 & 24.2$\times$ & 15.9$\times$ & $> 129.4\times$ \\

\midrule
\textbf{Avg.}
& \textbf{4272.35} & \textbf{6.52} & \textbf{--}
& \textbf{128.19}  & \textbf{0.30} & \textbf{0.065}
& \textbf{33.3$\times$} & \textbf{21.7$\times$} & \textbf{$>$61.6$\times$} \\

\bottomrule
\end{tabular}
}
\vspace{-2mm}
\end{table}

\begin{table}[htbp]
\centering
\footnotesize 
\caption{Case study on local subgraph optimization (ALS Effect) for \texttt{adder}, \texttt{multiplier}, and \texttt{arbiter} benchmarks ($\epsilon=10\%$). (units: Delay (ps), Area ($\mu m^2$)).}
\label{tab:subgraph_case_study_combined}
\setlength{\tabcolsep}{2pt} 
\renewcommand{\arraystretch}{1.1} 
\resizebox{\columnwidth}{!}{
\begin{tabular}{c c c cc cc c cc}
\toprule
\multirow{2}{*}{\textbf{Benchmark}} 
& \multirow{2}{*}{\textbf{Case}} 
& \multirow{2}{*}{\textbf{\#I/O}}
& \multicolumn{2}{c}{\textbf{Baseline}} 
& \multicolumn{2}{c}{\textbf{ALSRAC \cite{meng2020alsrac}}} 
& \textbf{CT \cite{li2025circuit}} 
& \multicolumn{2}{c}{\textbf{\MethodName{} (Ours)}} \\

\cmidrule(lr){4-5}
\cmidrule(lr){6-7}
\cmidrule(lr){8-8}
\cmidrule(lr){9-10}

& & & Delay & Area 
& Delay$\downarrow$ & Area$\downarrow$ 
& Delay/Area 
& Delay$\downarrow$ & Area$\downarrow$ \\
\midrule

\multirow{3}{*}{\texttt{adder}}
& Case 1 & 18/10 & 159.72 & 3.08 & 139.42 & 2.99 & OOM & \textbf{139.42} & \textbf{2.84} \\
& Case 2 & 23/17 & 128.99 & 3.41 & 128.99 & 3.39 & OOM & \textbf{128.99} & 3.41 \\
& Case 3 & 11/9  & 69.53  & 1.47 & 69.31 & 1.73 & OOM & \textbf{65.29} & 1.67 \\

\midrule

\multirow{5}{*}{\texttt{multiplier}} 
& Case 1 & 23/11 & 127.10 & 2.95 & 116.17 & 3.31 & OOM & \textbf{106.50} & \textbf{2.26} \\
& Case 2 & 28/13 & 63.23 & 2.93 & 64.80 & 2.14 & OOM & 64.80 & \textbf{2.13} \\
& Case 3 & 22/8  & 101.54 & 2.40 & 91.83 & 2.05 & OOM & 95.04 & 2.42 \\
& Case 4 & 28/15 & 196.18 & 3.90 & 186.67 & 3.57 & OOM & \textbf{152.21} & \textbf{3.09} \\
& Case 5 & 27/15 & 159.62 & 4.48 & 156.42 & 3.48 & OOM & \textbf{143.89} & 3.77 \\

\midrule

\multirow{5}{*}{\texttt{arbiter}} 
& Case 1 & 47/7 & 344.80 & 2.79 & 40.60 & 0.79 & OOM & \textbf{32.38} & \textbf{0.74} \\
& Case 2 & 51/2 & 392.61 & 2.63 & 32.38 & 0.21 & OOM & \textbf{31.83} & \textbf{0.16} \\
& Case 3 & 42/5 & 298.96 & 2.23 & 55.74 & 0.63 & OOM & \textbf{38.07} & \textbf{0.36} \\
& Case 4 & 53/3 & 274.35 & 2.78 & 55.74 & 0.87 & OOM & \textbf{38.76} & \textbf{0.72} \\
& Case 5 & 51/1 & 406.04 & 2.71 & 32.38 & 0.15 & OOM & \textbf{31.83} & \textbf{0.10} \\

\bottomrule
\end{tabular}
}
\vspace{-12pt}
\end{table}

\subsection{End-to-End Scalability on Large-Scale Circuits}

\wadd{Moving beyond localized subgraphs, we evaluate \MethodName{}'s global scalability on large EPFL circuits.
This stress-tests our entire partition-and-merge pipeline described in Sec.~\ref{subsec:scalable_framework}. Table~\ref{tab:ppa_epfl} reports the global PPA improvements under a 10\% ER bound. 

Notably, the baseline generative model, CT, suffers systemic OOM failures on these large designs, exposing its fundamental scalability limits. Furthermore, while heavily optimized heuristics like AppResub achieve deep PPA reductions via exhaustive incremental rewriting, they often incur prohibitive computational overhead when scaling to complex graphs. 

In contrast, \MethodName{} efficiently decomposes massive DAGs, optimizes subgraphs using the core engine, and seamlessly reinserts them. Overall, \MethodName{} achieves highly competitive average area and delay reductions of 25.18\% and 11.83\%, respectively. By approaching SOTA heuristic performance\gdel{ while drastically accelerating inference times}, \MethodName{} establishes a highly efficient, scalable generative paradigm for industry-grade circuits.}








\begin{table}[t!]
\centering
\footnotesize
\caption{End-to-End PPA Comparison on EPFL ($\epsilon=10\%$). (Units: Delay (ps), Area ($\mu m^2$)).}
\label{tab:ppa_epfl}

\setlength{\tabcolsep}{3pt}
\renewcommand{\arraystretch}{1.1}

\resizebox{\columnwidth}{!}{
\begin{tabular}{l rr rr rr c rr}
\toprule
\multirow{2}{*}{\textbf{Bench.}} 
& \multicolumn{2}{c}{\textbf{Baseline}} 
& \multicolumn{2}{c}{\textbf{ALSRAC \cite{meng2020alsrac}}}
& \multicolumn{2}{c}{\textbf{AppResub \cite{AppResub}}} 
& \textbf{CT \cite{li2025circuit}} 
& \multicolumn{2}{c}{\textbf{\MethodName{} (Ours)}} \\

\cmidrule(lr){2-3}
\cmidrule(lr){4-5}
\cmidrule(lr){6-7}
\cmidrule(lr){8-8}
\cmidrule(lr){9-10}

& Delay & Area 
& Delay$\downarrow$ & Area$\downarrow$
& Delay$\downarrow$ & Area$\downarrow$  
& Delay/Area 
& Delay$\downarrow$ & Area$\downarrow$  \\
\midrule

adder      & 3565.62 & 52.25   & 3565.62 & 52.25 & 3565.62 & 52.25  & OOM & 3565.62 & 52.25  \\
arbiter    & 1576.67 & 589.40  & 85.77   & 47.30 & 85.23   & 47.73   & OOM & 1174.48 & 160.41 \\
calvc      & 346.08  & 25.66   & 214.59  & 10.64 & 219.07  & 12.95   & OOM & 324.65  & 23.33  \\
ctrl       & 146.23  & 4.86    & 150.55  & 4.22  & 122.34  & 3.97    & OOM & 148.54  & 4.79   \\
i2c        & 303.29  & 47.67   & 256.26  & 19.76 & 256.91  & 19.81   & OOM & 303.29  & 45.95  \\
multiplier & 7112.19 & 1041.10 & 6558.47 & 1040.77 & 7110.63 & 1037.86 & OOM & 7102.56 & 1040.80\\
priority   & 2715.86 & 32.19   & 18.47   & 0.26  & 19.64   & 0.22    & OOM & 1281.60 & 14.11  \\

\midrule
\textbf{Avg.} & \textbf{2252.28} & \textbf{256.16} & \textbf{1549.96} & \textbf{167.89} & \textbf{1625.63} & \textbf{167.83} & \textbf{OOM} & \textbf{1985.82} & \textbf{191.66} \\
\bottomrule
\end{tabular}
}
\vspace{-10pt}
\end{table}

\section{Conclusion and Future Work}\label{sect_concl}

In this paper, we introduced \MethodName{}, an end-to-end framework designed to overcome the severe scalability bottlenecks of generative circuit design. 
By synergizing a scalable partition-and-merge pipeline with a specialized Transformer core—empowered by irredundant encoding and a masking mechanism—\MethodName{} successfully extends generative approximate logic synthesis to arbitrary-scale designs. 
It establishes a new paradigm by consistently discovering superior Pareto-optimal PPA trade-offs compared to traditional ALS heuristics and exact generative baselines. 
Future research will explore advanced graph representations to enable Transformers to process massive circuits directly, further unleashing the potential of generative EDA.

\balance
\bibliographystyle{unsrtnat} %
\bibliography{main}

\end{document}